\documentclass[11pt]{article}
\usepackage{graphicx}

\title{A Four Function Variational Principle for Barotropic Magnetohydrodynamics}
\author{Asher Yahalom$^{a}$ \\
$^a$ Ariel University Center of Samaria, Ariel 40700, Israel\\
e-mail: asya@ariel.ac.il; }

\begin{document}
\maketitle

\newcommand{\beq} {\begin{equation}}
\newcommand{\enq} {\end{equation}}
\newcommand{\ber} {\begin {eqnarray}}
\newcommand{\enr} {\end {eqnarray}}
\newcommand{\eq} {equation}
\newcommand{\eqn} {equation }
\newcommand{\eqs} {equations }
\newcommand{\ens} {equations}
\newcommand{\mn}  {{\mu \nu}}
\newcommand {\er}[1] {equation (\ref{#1}) }
\newcommand {\ern}[1] {equation (\ref{#1})}
\newcommand {\ers}[1] {equations (\ref{#1})}
\newcommand {\Er}[1] {Equation (\ref{#1}) }
\newcommand {\ncite}[1] {[\cite{#1}] }

\begin{abstract}
Variational principles for magnetohydrodynamics were introduced by
previous authors both in Lagrangian and Eulerian form. In a
previous work Yahalom \& Lynden-Bell introduced a simpler Eulerian
variational principles from which all the relevant
 equations of magnetohydrodynamics can be derived.
 The variational principle was given in terms of six independent functions
for non-stationary flows and three independent functions for
stationary flows. This is less then the seven variables which
appear in the standard equations of magnetohydrodynamics which are
the magnetic field $\vec B$ the velocity field $\vec v$ and the
density $\rho$. In this work I will improve on the previous
results showing that non-stationary magnetohydrodynamics should be
described by four functions .
\end{abstract}

\section{Introduction}

Variational principles for
magnetohydrodynamics were introduced by previous authors both in
Lagrangian and Eulerian form. Sturrock \cite{Sturrock} has
discussed in his book a Lagrangian variational formalism for
magnetohydrodynamics. Vladimirov and Moffatt \cite{Moffatt} in a
series of papers have discussed an Eulerian variational principle
for incompressible magnetohydrodynamics. However, their
variational principle contained three more functions in addition
to the seven variables which appear in the standard equations of
magnetohydrodynamics which are the magnetic field $\vec B$ the
velocity field $\vec v$ and the density $\rho$. Kats \cite{Kats}
has generalized Moffatt's work for compressible non barotropic
flows but without reducing the number of functions and the
computational load. Moreover, Kats have shown that the variables
he suggested can be utilized to describe the motion of arbitrary
discontinuity surfaces \cite{Kats3,Kats4}. Sakurai \cite{Sakurai}
has introduced a two function Eulerian variational principle for
force-free magnetohydrodynamics and used it as a basis of a
numerical scheme, his method is discussed in a book by Sturrock
\cite{Sturrock}. A method of solving the equations for those two
variables was introduced by Yang, Sturrock \& Antiochos
\cite{Yang}. In a
 recent work Yahalom \& Lynden-Bell \cite{YaLy,Yahalom2} have
 combined the Lagrangian of Sturrock \cite{Sturrock}
with the Lagrangian of Sakurai \cite{Sakurai} to obtain an {\bf
Eulerian} Lagrangian principle depending on only six
functions. The vanishing of the variational derivatives of this Lagrangian entail
 all the equations needed to describe barotropic
magnetohydrodynamics without any additional constraints. The
equations obtained resemble the equations of Frenkel, Levich \&
Stilman \cite{FLS} (see also \cite{Zakharov}). Furthermore, it was
shown that for stationary flows three functions will suffice in
order to describe a Lagrangian principle for barotropic
magnetohydrodynamics. The non-singlevaluedness of the functions
appearing in the reduced representation of barotropic
magnetohydrodynamics was discussed in particular with connection
to the topological invariants of magnetic and cross helicities. It
was shown how the conservation of cross helicity can be easily
generated using the Noether theorem and the variables introduced in that paper.
In the current paper I improve on the previous results and show that four functions are enough to describe
a general non stationary barotropic magnetohydrodynamics, the idea is borrowed from \cite{YaLy2} see also \cite{Yahalom3,Yahalom4,Yahalom5}.

The plan of this paper is as follows: First I introduce the
standard notations and equations of barotropic
magnetohydrodynamics. Next I introduce the potential
representation of the magnetic field $\vec B$ and the velocity
field $\vec v$. This is followed by a review of the Eulerian
variational principle developed by  Yahalom \& Lynden-Bell
\cite{YaLy,Yahalom2}. After those introductory sections I will
present the four function Eulerian variational principles for
non-stationary magnetohydrodynamics.

\section{The standard formulation of barotropic magnetohydrodynamics}

The standard set of \eqs solved for barotropic magnetohydrodynamics are given below:
\beq
\frac{\partial{\vec B}}{\partial t} = \vec \nabla \times (\vec v \times \vec B),
\label{Beq}
\enq
\beq
\vec \nabla \cdot \vec B =0,
\label{Bcon}
\enq
\beq
\frac{\partial{\rho}}{\partial t} + \vec \nabla \cdot (\rho \vec v ) = 0,
\label{masscon}
\enq
\beq
\rho \frac{d \vec v}{d t}=
\rho (\frac{\partial \vec v}{\partial t}+(\vec v \cdot \vec \nabla)\vec v)  = -\vec \nabla p (\rho) +
\frac{(\vec \nabla \times \vec B) \times \vec B}{4 \pi}.
\label{Euler}
\enq
The following notations are utilized: $\frac{\partial}{\partial t}$ is the temporal derivative,
$\frac{d}{d t}$ is the temporal material derivative and $\vec \nabla$ has its
standard meaning in vector calculus. $\vec B$ is the magnetic field vector, $\vec v$ is the
velocity field vector and $\rho$ is the fluid density. Finally $p (\rho)$ is the pressure which
we assume depends on the density alone (barotropic case). The justification for those \eqs
and the conditions under which they apply can be
found in standard books on magnetohydrodynamics (see for example \cite{Sturrock}). \Er{Beq}describes the
fact that the magnetic field lines are moving with the fluid elements ("frozen" magnetic field lines),
 \ern{Bcon} describes the fact that
the magnetic field is solenoidal, \ern{masscon} describes the conservation of mass and \ern{Euler}
is the Euler equation for a fluid in which both pressure
and Lorentz magnetic forces apply. The term:
\beq
\vec J =\frac{\vec \nabla \times \vec B}{4 \pi},
\label{J}
\enq
is the electric current density which is not connected to any mass flow.
The number of independent variables for which one needs to solve is seven
($\vec v,\vec B,\rho$) and the number of \eqs (\ref{Beq},\ref{masscon},\ref{Euler}) is also seven.
Notice that \ern{Bcon} is a condition on the initial $\vec B$ field and is satisfied automatically for
any other time due to \ern{Beq}. Also notice that $p (\rho)$ is not a variable rather it is a given
function of $\rho$.

\section{Potential representation of vector quantities of magnetohydrodynamics}

It was shown in \cite{YaLy} that $\vec B$ and $\vec v$ can be represented in terms of five scalar functions
$\alpha,\beta,\chi,\eta,\nu$. Following Sakurai \cite{Sakurai}
 the magnetic field takes the form:
\beq
\vec B = \vec \nabla \chi \times \vec \nabla \eta.
\label{Bsakurai}
\enq
Hence $\vec B$ satisfies automatically \er{Bcon} and is orthogonal to both $\vec \nabla \chi$ and $\vec \nabla \eta$.
A similar representation was suggested by Dungey \cite{Dungey} but not in the context of variational analysis.
The above expression can also describe a magnetic field with non-zero magnetic helicity as was demonstrated in \cite{YaLy}.
Moreover, the velocity $\vec v$ can be represented in the following form:
\beq
\vec v =  \vec \nabla \nu + \alpha \vec \nabla \chi + \beta \vec \nabla \eta.
\label{vform}
\enq
this representation is a generalization of the Clebsch representation \cite{Lamb H.} for magnetohydrodynamics.

\section{The Action of Barotropic Magnetohydrodynamics}

It was shown in \cite{YaLy} that the action of barotropic magnetohydrodynamics takes the form:
\ber
A & \equiv & \int {\cal L} d^3 x dt,
\nonumber \\
 {\cal L} &\equiv & -\rho \left[\frac{\partial{\nu}}{\partial t} + \alpha \frac{\partial{\chi}}{\partial t}
+ \beta \frac{\partial{\eta}}{\partial t}+\varepsilon (\rho)+
\frac{1}{2} (\vec \nabla \nu + \alpha \vec \nabla \chi +  \beta \vec \nabla \eta)^2 \right]
\nonumber \\
&-&\frac{1}{8 \pi}(\vec \nabla \chi \times \vec \nabla \eta)^2,
\label{Lagactionsimp}
\enr
in which $\varepsilon (\rho)$ is the specific internal energy.
Taking the variational derivatives to zero for arbitrary variations leads to the following set of equations:
\beq
\frac{\partial{\rho}}{\partial t} + \vec \nabla \cdot (\rho \vec v ) = 0,
\label{masscon2}
\enq
\beq
 \frac{d \chi}{dt} = 0,
 \label{chieq}
\enq
\beq
  \frac{d \eta}{dt} = 0,
\label{etaeq}
\enq
\beq
\frac{d \nu}{d t} = \frac{1}{2} \vec v^2 - w,
\label{nueq}
\enq
in which $w$ is the specific enthalpy.
\beq
\frac{d \alpha}{dt} = \frac{\vec \nabla \eta \cdot \vec J}{\rho}, \qquad
\label{aleq}
\enq
\beq
\frac{d \beta}{dt} = -\frac{\vec \nabla \chi \cdot \vec J}{\rho}.
\label{betaeq}
\enq
In all the above equations  $\vec B$ is given by \er{Bsakurai} and $\vec v$ is given by \er{vform}.
The mass conservation  \ern{masscon} is readily obtained. Now one needs to show
that also \er{Beq} and \er{Euler} are satisfied.

It can be easily shown that provided that $\vec B$ is in the form given in \ern{Bsakurai},
and \ern{chieq} and \ern{etaeq} are satisfied, then \ers{Beq}  are satisfied.

We shall now show that a velocity field given by \ern{vform}, such that the
\eqs for $\alpha, \beta, \chi, \eta, \nu$ satisfy the corresponding equations
(\ref{masscon2},\ref{chieq},\ref{etaeq},\ref{nueq},\\
\ref{aleq},\ref{betaeq}) must satisfy Euler's equations.
Let us calculate the material derivative of $\vec v$:
\beq
\frac{d\vec v}{dt} = \frac{d\vec \nabla \nu}{dt}  + \frac{d\alpha}{dt} \vec \nabla \chi +
 \alpha \frac{d\vec \nabla \chi}{dt}  +
\frac{d\beta}{dt} \vec \nabla \eta + \beta \frac{d\vec \nabla \eta}{dt}.
\label{dvform}
\enq
It can be easily shown that:
\ber
\frac{d\vec \nabla \nu}{dt} & = & \vec \nabla \frac{d \nu}{dt}- \vec \nabla v_k \frac{\partial \nu}{\partial x_k}
 = \vec \nabla (\frac{1}{2} \vec v^2 - w)- \vec \nabla v_k \frac{\partial \nu}{\partial x_k},
 \nonumber \\
 \frac{d\vec \nabla \eta}{dt} & = & \vec \nabla \frac{d \eta}{dt}- \vec \nabla v_k \frac{\partial \eta}{\partial x_k}
 = - \vec \nabla v_k \frac{\partial \eta}{\partial x_k},
 \nonumber \\
 \frac{d\vec \nabla \chi}{dt} & = & \vec \nabla \frac{d \chi}{dt}- \vec \nabla v_k \frac{\partial \chi}{\partial x_k}
 = - \vec \nabla v_k \frac{\partial \chi}{\partial x_k}.
 \label{dnabla}
\enr
In which $x_k$ is a Cartesian coordinate and a summation convention is assumed. Equations (\ref{chieq},\ref{etaeq},\ref{nueq})
where used in the above derivation. Inserting the result from equations (\ref{aleq},\ref{betaeq},\ref{dnabla})
into \ern{dvform} yields:
\ber
\frac{d\vec v}{dt} &=& - \vec \nabla v_k (\frac{\partial \nu}{\partial x_k} + \alpha \frac{\partial \chi}{\partial x_k} +
\beta \frac{\partial \eta}{\partial x_k}) + \vec \nabla (\frac{1}{2} \vec v^2 - w)
 \nonumber \\
&+& \frac{1}{\rho} ((\vec \nabla \eta \cdot \vec J)\vec \nabla \chi - (\vec \nabla \chi \cdot \vec J)\vec \nabla \eta)
 \nonumber \\
&=& - \vec \nabla v_k v_k + \vec \nabla (\frac{1}{2} \vec v^2 - w)
 + \frac{1}{\rho} \vec J \times (\vec \nabla \chi \times  \vec \nabla \eta)
 \nonumber \\
&=& - \frac{\vec \nabla p}{\rho} + \frac{1}{\rho} \vec J \times \vec B.
\label{dvform2}
\enr
In which we have used both \ern{Bsakurai}  and \ern{vform} in the above derivation. This of course
proves that the barotropic Euler equations can be derived from the action given in \er{Lagactionsimp} and hence
all the equations of barotropic magnetohydrodynamics can be derived from the above action
without restricting the variations in any way except on the relevant boundaries and cuts.
The reader should take into account that the topology of the magnetohydrodynamic flow is conserved,
hence cuts must be introduced into the calculation as initial conditions.

\section{A Simpler Action for Barotropic Magnetohydrodynamics}

Can we obtain a further reduction of barotropic magnetohydrodynamics? Can we
formulate magnetohydrodynamics with less than the six functions $\alpha,\beta,\chi,\eta,\nu,\rho$?
The answer is yes, in fact four functions $\chi,\eta,\nu,\rho$ will suffice.
To see this we may write the two \eqs (\ref{chieq},\ref{etaeq}) as \eqs for $\alpha,\beta$ that is:
\ber
& &  \frac{d \chi}{dt} = \frac{\partial \chi}{\partial t}+ \vec v \cdot \vec \nabla \chi  =
\frac{\partial \chi}{\partial t}+
(\vec \nabla \nu + \alpha \vec \nabla \chi + \beta \vec \nabla \eta) \cdot \vec \nabla \chi = 0,
\nonumber \\
& &  \frac{d \eta}{dt} = \frac{\partial \eta}{\partial t}+ \vec v \cdot \vec \nabla \eta  =
\frac{\partial \eta}{\partial t}+
(\vec \nabla \nu + \alpha \vec \nabla \chi + \beta \vec \nabla \eta) \cdot \vec \nabla \eta = 0,
\label{lagmul2}
\enr
in which we have used \ern{vform}. Solving for $\alpha,\beta$ we obtain:
\ber
\alpha[\chi,\eta,\nu] & = &  \frac{(\vec \nabla \eta)^2(\frac{\partial \chi}{\partial t}+ \vec \nabla \nu \cdot \vec \nabla \chi)
- (\vec \nabla \eta \cdot \vec \nabla \chi) (\frac{\partial \eta}{\partial t}+ \vec \nabla \nu \cdot \vec \nabla \eta)}
{(\vec \nabla \eta \cdot \vec \nabla \chi)^2-(\vec \nabla \eta)^2 ( \vec \nabla \chi)^2 }
\nonumber \\
\beta[\chi,\eta,\nu] & = & \frac{(\vec \nabla \chi)^2(\frac{\partial \eta}{\partial t}+ \vec \nabla \nu \cdot \vec \nabla \eta)
- (\vec \nabla \eta \cdot \vec \nabla \chi) (\frac{\partial \chi}{\partial t}+ \vec \nabla \nu \cdot \vec \nabla \chi)}
{(\vec \nabla \eta \cdot \vec \nabla \chi)^2-(\vec \nabla \eta)^2 ( \vec \nabla \chi)^2 }.
\label{alphbeta}
\enr
Hence $\alpha$ and $\beta$ are not free variables any more, but depend on $\chi,\eta,\nu$. Moreover,
the velocity $\vec v$ now depends on the same three variables $\chi,\eta,\nu$:
\beq
\vec v =  \vec \nabla \nu + \alpha[\chi,\eta,\nu] \vec \nabla \chi + \beta[\chi,\eta,\nu] \vec \nabla \eta.
\label{vform2}
\enq
Since $\vec v$ is given now  by \ern{vform2} it follows that the two \eqs (\ref{chieq},\ref{etaeq}) are satisfied identically
and need not be derived from a variational principle. The above equation can be somewhat simplified resulting in:
\ber
\vec v  &=&  \vec \nabla \nu + \frac{1}{\vec B^2} [\frac{\partial \eta}{\partial t} \vec \nabla \chi  -
\frac{\partial \chi}{\partial t} \vec \nabla \eta + \vec \nabla \nu \times \vec B]\times \vec B
\nonumber \\
&=& \frac{1}{\vec B^2} [(\frac{\partial \eta}{\partial t} \vec \nabla \chi  -
\frac{\partial \chi}{\partial t} \vec \nabla \eta) \times \vec B + \vec B (\vec \nabla \nu \cdot \vec B)]
\label{vform3}
\enr
Hence the velocity $\vec v$ is partitioned naturally into two components one which is parallel to the magnetic field
and another one which is perpendicular to it:
\ber
\vec v  &=&   \vec v_{\bot}+ \vec v_{\|}
\nonumber \\
\vec v_{\bot} &=& \frac{1}{\vec B^2} (\frac{\partial \eta}{\partial t} \vec \nabla \chi  -
\frac{\partial \chi}{\partial t} \vec \nabla \eta) \times \vec B,
\qquad
\vec v_{\|} = \frac{\vec B}{\vec B^2}  (\vec \nabla \nu \cdot \vec B).
\label{vform4}
\enr
Inserting the velocity representation (\ref{vform3}) into \ern{alphbeta} will lead to the result:
\ber
\alpha & = &   \frac{\vec \nabla \eta \cdot (\vec B \times (\vec v- \vec \nabla \nu))}{\vec B^2}
\nonumber \\
\beta  & = &  - \frac{\vec \nabla \chi \cdot (\vec B \times (\vec v- \vec \nabla \nu))}{\vec B^2}.
\label{alphbeta2}
\enr
Finally \ers{alphbeta} should be substituted into
\ern{Lagactionsimp} to obtain a Lagrangian density ${\cal L}$ in
terms of $\chi,\eta,\nu,\rho$.
\ber
 {\cal L}[\chi,\eta,\nu,\rho] &\equiv & -\rho [\frac{\partial{\nu}}{\partial t} + \alpha[\chi,\eta,\nu] \frac{\partial{\chi}}{\partial t}
+ \beta[\chi,\eta,\nu] \frac{\partial{\eta}}{\partial t}+\varepsilon (\rho)
\nonumber \\
&+& \frac{1}{2} (\vec \nabla \nu + \alpha[\chi,\eta,\nu] \vec \nabla \chi +  \beta[\chi,\eta,\nu] \vec \nabla \eta)^2 ]
\nonumber \\
&-&\frac{1}{8 \pi}(\vec \nabla \chi \times \vec \nabla \eta)^2.
\label{Lagsimp2}
\enr
Using \ers{alphbeta2} this can be written as:
\beq
 {\cal L}[\chi,\eta,\nu,\rho] = \rho [\frac{1}{2} \vec v^2 - \frac{d{\nu}}{d t}-\varepsilon (\rho)] -  \frac{1}{8 \pi}\vec B^2
\label{Lagsimp3}
 \enq
were $\vec v$ is given by \ern{vform3} and $\vec B$ by \ern{Bsakurai}. Or more explicitly as:
\ber
 {\cal L}[\chi,\eta,\nu,\rho] &=& \frac{1}{2} \frac{\rho}{(\vec \nabla \chi \times \vec \nabla \eta)^2}
  [\vec \nabla \eta \frac{\partial \chi}{\partial t}- \vec \nabla \chi \frac{\partial \eta}{\partial t}+
  (\vec \nabla \chi \times \vec \nabla \eta) \times \vec \nabla \nu]^2
  \nonumber \\   &-& \rho [\frac{\partial \nu}{\partial t} + \frac{1}{2} (\vec \nabla \nu)^2 + \varepsilon (\rho)]
-  \frac{(\vec \nabla \chi \times \vec \nabla \eta)^2}{8 \pi}.
\label{Lagsimp4}
 \enr
This Lagrangian density admits an infinite symmetry group of transformations of the form:
\beq
\hat{\eta} =  \hat{\eta} (\chi,\eta), \qquad \hat{\chi} =  \hat{\chi} (\chi,\eta),
\enq
provided that the absolute value of the Jacobian of these transformation is unity:
\beq
\left|\frac{\partial (\hat{\eta},\hat{\chi})}{\partial (\eta,\chi)}\right|=1.
\enq
In particular the Lagrangian density admits an exchange symmetry:
\beq
\hat{\eta} =  \chi, \qquad \hat{\chi} = \eta.
\enq
As a consequence of the double infinite symmetry group we have two {\it local} conservation laws
given by the two \eqs (\ref{chieq},\ref{etaeq}). Taking the variational derivatives of the action defined using \ern{Lagsimp4}
to zero for arbitrary variations leads to the following set of equations:
\beq
\frac{\partial{\rho}}{\partial t} + \vec \nabla \cdot (\rho \vec v ) = 0,
\label{masscon3}
\enq
\beq
\frac{d \nu}{d t} = \frac{1}{2} \vec v^2 - w,
\label{nueq2}
\enq
\beq
\frac{d \alpha[\chi,\eta,\nu]}{dt} = \frac{\vec \nabla \eta \cdot \vec J}{\rho}, \qquad
\label{aleq2}
\enq
\beq
\frac{d \beta[\chi,\eta,\nu]}{dt} = -\frac{\vec \nabla \chi \cdot \vec J}{\rho}.
\label{betaeq2}
\enq
Those equations should be solved for $\chi,\eta,\nu,\rho$. Equations (\ref{aleq2},\ref{betaeq2}) contain a complicated linear
combination of the second derivatives $\frac{\partial^2 \chi}{\partial t^2}$ and $\frac{\partial^2 \eta}{\partial t^2}$.
This is inconvenient numerically therefore the following approach is recommended. Taking the partial temporal derivative of
the two \eqs (\ref{chieq},\ref{etaeq}) we obtain:
\beq
\frac{\partial^2 \chi}{\partial t^2}+ \frac{\partial \vec v}{\partial t} \cdot \vec \nabla \chi +
 \vec v \cdot \vec \nabla \frac{\partial \chi}{\partial t} = 0, \qquad
 \frac{\partial^2 \eta}{\partial t^2}+ \frac{\partial \vec v}{\partial t} \cdot \vec \nabla \eta +
 \vec v \cdot \vec \nabla \frac{\partial \eta}{\partial t} = 0.
 \label{d2chid2eta1}
 \enq
Using the expression $\frac{\partial \vec v}{\partial t}$ from \ern{dvform2} we obtain an explicit expression for the second derivatives
of the form:
\ber
\frac{\partial^2 \chi}{\partial t^2}&=&((\vec v \cdot \vec \nabla) \vec v+ \vec \nabla w -\frac{1}{\rho} \vec J \times \vec B )\cdot \vec \nabla \chi
-\vec v \cdot \vec \nabla \frac{\partial \chi}{\partial t}
\nonumber \\
\frac{\partial^2 \eta}{\partial t^2}&=&((\vec v \cdot \vec \nabla) \vec v+ \vec \nabla w -\frac{1}{\rho} \vec J \times \vec B )\cdot \vec \nabla \eta
-\vec v \cdot \vec \nabla \frac{\partial \eta}{\partial t}.
\label{d2chid2eta2}
 \enr
Hence we arrived at a four function formalism for barotropic magnetohydrodynamics which can be derived from a Lagrangian. Notice,
however, that this formalism contains two first order equations and two second order equations, while our previous six function
formalism \cite{YaLy} contained six first order equations.

\section{Conclusion}

We have shown that barotropic magnetohydrodynamics can be represented in terms of four scalar functions
$\chi,\eta,\nu,\rho$ instead of the seven quantities which are the magnetic field $\vec B$ the velocity field $\vec v$ and the
density $\rho$. Anticipated applications include stability analysis and the description of
numerical schemes using the described variational principles, exceed the scope of this paper.

It was shown by the author \cite{Yahalom} that variational principles can be used directly for numerical analysis
(simulation) without the need to refer to the field equations. This mathematical construction may lead to better algorithms for
simulating magnetohydrodynamics in terms of the needed computer memory and CPU time. This approach was applied to potential flows in a series
of papers \cite{YahalomPinhasi,YahPinhasKop,OphirYahPinhasKop}. Moreover, it was implemented in a user friendly software
 package FLUIDEX (which can be down loaded from the web site www.fluidex-cfd.com). A variational formalism of
 magnetohydrodynamics should serve the same use.

As for stability analysis I suspect that for achieving this we will need to add additional
constants of motion constraints to the action as was done by  \cite{YahalomKatz}, hopefully this will be discussed in a future paper.

\end{document}